\documentstyle[11pt,moriond,epsfig]{article}

\newcommand \beq{\begin{eqnarray}}
\newcommand \eeq{\end{eqnarray}}
\newcommand{\be}{\begin{eqnarray}}
\newcommand{\ee}{\end{eqnarray}}
\newcommand{\nn}{\nonumber\\ }

\def\del{\partial}                              

\def\frac#1#2{{#1 \over #2}}
\def\simge{\mathrel{%
   \rlap{\raise 0.511ex \hbox{$>$}}{\lower 0.511ex \hbox{$\sim$}}}}
\def\simle{\mathrel{
   \rlap{\raise 0.511ex \hbox{$<$}}{\lower 0.511ex \hbox{$\sim$}}}}

\def\be{\begin{equation}}
\def\ee{\end{equation}}
\def\bea{\begin{eqnarray}}
\def\eea{\end{eqnarray}}

\begin{document}

\begin{titlepage}
\begin{flushright} {CERN-TH/2000-142\\hep-ph/0005195}
\end{flushright}
\vspace*{1.5cm}

\vspace*{4cm}
\title{SMALL x,  SATURATION AND THE HIGH ENERGY LIMIT OF QCD}

\author{Edmond IANCU}

\address{CERN, Theory Division \\ CH-1211 Geneva 23, Switzerland}

\maketitle\abstracts{The high energy limit of QCD is controlled 
by the small-$x$ part of a hadron
wavefunction.  We argue that this part is universal to
all hadrons and is composed of a new form of matter: a Colored Glass
Condensate.  This matter is weakly interacting at very small $x$, but is
non-perturbative because of the highly occupied boson states which compose
the condensate.  Such a matter might be studied in high energy lepton-hadron
or hadron-hadron interactions.}

\vspace*{7.5cm}

\begin{flushleft}
Invited talk given at the XXXVth Rencontres
de Moriond \\
QCD and Hadronic Interactions\\
March 18 -- 25, 2000, Les Arcs, France
\end{flushleft}


\end{titlepage}

\newpage 
\vspace*{4cm}
\title{SMALL x,  SATURATION AND THE HIGH ENERGY LIMIT OF QCD \footnote{
Work done in collaboration with Andrei Leonidov and
Larry McLerran \cite{CGC}.} }

\author{Edmond IANCU}

\address{CERN, Theory Division \\ CH-1211 Geneva 23, Switzerland}

\maketitle\abstracts{The high energy limit of QCD is controlled 
by the small-$x$ part of a hadron
wavefunction.  We argue that this part is universal to
all hadrons and is composed of a new form of matter: a Colored Glass
Condensate.  This matter is weakly interacting at very small $x$, but is
non-perturbative because of the highly occupied boson states which compose
the condensate.  Such a matter might be studied in high energy lepton-hadron
or hadron-hadron interactions.}

\section{The Colored Glass Condensate}

At very small values of the Bjorken $x$-variable, one expects
QCD to enter a new regime which is caracterized by parton saturation
and very high values of the QCD field strength $F^a_{\mu\nu}\sim 1/g$.
Saturation, which is a limitation on the maximum phase-space parton 
density that can be reached in the hadron wavefunction, may have been
already observed at HERA \cite{CALD}, and should play an important
role in the very early stages of
relativistic heavy ion collisions at RHIC and LHC \cite{AM1}
(and Refs. therein). 
In the saturation regime, the individual parton-parton interactions
may be weak \footnote{This is the case if the saturation
momentum $Q_s$ (cf. eq.~(\ref{SATN}) below) is large enough; e.g., at LHC,
one expects $Q_s\sim 2-3$ GeV.} (which we shall assume in what follows;
i.e., we assume that $g\ll 1$), but the parton densities are so
large that the system becomes strongly non-perturbative. Thus, at
a theoretical level, understanding saturation is a challenging and 
fascinating problem where one has to deal with fully non-linear QCD.
This is reminiscent of a similar problem in high temperature QCD
where perturbation theory breaks down at the soft scale $g^2T$
because of the large thermal occupation numbers of the soft gluons
\cite{SEWM98} (and Refs. therein).

The efforts toward understanding the region of high gluon density
have uncovered a new form of matter which is formed from these gluons 
\cite{CGC}. This matter is universal in that it should describe the high
gluon density part of any hadron and nuclear
wavefunction at small $x$. The combination 
of high density and the fact that gluons are massless bosons leads 
naturally to the expectation that this matter is a Bose condensate.  
Since the gluons carry color and local color is a gauge dependent quantity, 
any gauge invariant formulation will neccessarily involve an average over 
all colors to restore the invariance.  This averaging procedure bears a 
formal ressemblance to the averaging over background fields done for spin 
glasses \cite{Parisi}.
The matter is therefore called the Colored Glass Condensate (CGC).

It should be emphasized that this picture holds, strictly
speaking, only in the infinite-momentum frame, 
where the hadron propagates almost at the speed of light, and
thus appears as an infinitesimally thin two-dimensional sheet
(by Lorentz contraction).  In this frame, the
parton interpretation makes sense and 
deep inelastic scattering (DIS) proceeds via the
instantaneous absorbtion of the external probe 
(e.g., a virtual photon $\gamma^*$ with 4-momentum $q^\mu$) 
by some parton in the hadron. The Bjorken $x_B$ parameter is defined as
$x_B\equiv Q^2/2P\cdot q$, where $Q^2 \equiv -q^\mu q_\mu$, and $P^\mu=
\delta^{\mu +}P^+$, with large $P^+$, is the hadron  4-momentum
\footnote{We use light-cone vector notations: for some arbitrary
vector $p^\mu$, we write $p^\mu=(p^+,p^-,{\bf p}_\perp)$, with
$p^+\equiv (1/\sqrt 2)(p^0+p^3)$,
$p^-\equiv (1/\sqrt 2)(p^0-p^3)$, and ${\bf p}_\perp
\equiv (p^1,p^2)$.}. By kinematics, $x_B$ coincides with
the longitudinal momentum fraction $x \equiv p^+/P^+$ of the
struck parton: $x_B = x$.

\begin{figure}
\protect\epsfxsize=5.cm{\centerline{\epsfbox{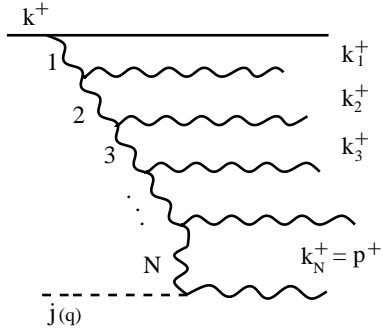}}}
         \caption{A parton cascade leading to a small-$x$ virtual
gluon.}
\label{Born}
\end{figure}
At $x\ll 1$, the
gluon density increases faster, and is the driving force toward 
saturation \footnote{To directly measure the gluon density, it is convenient to
consider a Gedanken experiment where the DIS is initiated by 
the ``current'' $j\equiv -\frac{1}{4}F^{\mu\nu}_aF^a_{\mu\nu}$
which couples directly to gluons.}. The dynamics that leads to this
increase is the quantum evolution toward small-$x$ : in a parton 
cascade initiated by some {\it fast} parton (i.e., a hadron constituent
with a relatively large longitudinal momentum $k^+\sim P^+$, e.g.,
a valence quark), the number
of radiated gluons increases exponentially with the rapidity gap
$\Delta \tau\equiv \ln(k^+/p^+) \sim \ln(1/x)$
between the original parton and the {\it soft} (i.e., small $p^+$:
$p^+=xP^+\ll P^+$) final gluon which is
struck by the external current (see Fig. 1).
This (BFKL) picture---which assumes the radiated gluons to behave as free
partons---ceases to be valid at very small $x$, where the gluon density 
is so large that the radiated gluons overlap each other in the 
transverse plane. This is the onset of saturation.

This is also the regime where the description in terms of 
a colored class condensate becomes appropriate: Corresponding
to the strong ordering in longitudinal momenta along the cascade,
\beq\label{HL}
k^+\equiv k_0^+\,\gg\,k_1^+\,\gg\,k_2^+\,\gg\,\cdots\,\gg\,
k_N^+\equiv p^+\,,\eeq
there is a similar hierarchy among the lifetimes of the
radiated gluons (since the latter are proportional to the
respective $k^+$ momenta):
Softer a gluon is, shorter is its lifetime.

This hierarchy in time stays at the basis of a {\it quenched
approximation} \cite{MV,JKMW97,JKLW1,JKLW2}
for the ``fast'' degrees of freedom: when 
``seen'' by the soft modes with very short lifetimes, the 
modes with larger $p^+$ appear as frozen (no dynamics)
and can be replaced by a {\it static} (i.e., independent
of $x^+$) and {\it random} color
charge configuration, with density $\rho_a(x^-,x_\perp)$.
(This is random since the soft gluons can belong to different
cascades.) The spatial correlations of the effective 
charge $\rho_a(x^-,x_\perp)$ reflect the quantum
dynamics at large longitudinal momenta, and are encoded in a statistical 
weight function  $W[\rho]$. 
Because of the Lorentz contraction, we can write
$\rho_a(x^-,x_\perp)\approx \delta(x^-) \rho_a(x_\perp)$,
and $W$ is a functional of the {\it superficial} charge density
$\rho_a(x_\perp)$ alone.

Thus, in order to compute soft correlations in this approximation,
one has to first study the (quantum) dynamics of the soft gluons 
in the presence of a given color charge $\rho_a$,
and then perform a (classical)
average over $\rho_a$, with weight function $W[\rho]$. 
Clearly, the latter will depend upon what we call ``soft'' and
``fast'', i.e., upon the separation scale $\Lambda$ 
between {\it fast} ($p^+>\Lambda$) and {\it soft} 
($p^+ <\Lambda$) degrees of freedom. We thus write
$W[\rho]\equiv W_\Lambda[\rho]$.

For instance, the 2-point correlation function
is obtained as (in the light-cone gauge $A^+=0$)
\be\label{2point}
\langle {\rm T}\,
A^\mu(x)A^\nu(y)\rangle
\,=\,\int {\cal D}\rho\,\,W_\Lambda[\rho]
\left\{\frac{\int^\Lambda {\cal D}A
\,\,A^\mu(x)A^\nu(y)\,\,{\rm e}^{\,iS[A,\,\rho]}}
{\int^\Lambda {\cal D}A\,\,{\rm e}^{\,iS[A,\,\rho]}}
\right\},\ee
with the functional integral running only over
soft gluon fields $A^\mu_a(p)$ with $p^+ <\Lambda$, and the
action $S[A,\rho]$ describing the dynamics of
these fields in the presence of the
classical source $\rho_a$, in the eikonal appoximation \cite{JKLW1}:  
$S=S_{YM}+S_W$, where
$S_{YM}=\int d^4x(-F_{\mu\nu}^2/4)$ is the usual Yang-Mills action, and
(with ${\tilde x}\equiv (x^-,{x}_{\perp})$, and T denoting
time-ordering of color matrices):
\be
S_W\,\equiv\,
{i \over {N_c}} \int d^3 \tilde x\, {\rm Tr}\,\left\{ \rho(\tilde x)
\,{\rm T}\, \exp\left[\,
ig\int dx^+ A^-(x^+,\tilde x) \right]\right\}\,.\ee
The double averaging in eq.~(\ref{2point}) is very similar to the one
performed for spin systems in a random external magnetic field \cite{Parisi} 
(which plays the role of  $\rho$ in the above equation), 
and supports the physical picture of the 
saturation regime as a coloured glass condensate.

Of course, the final results for soft correlators must be
independent of the arbitrary separation scale $\Lambda$.
The $\Lambda$-dependence of the weight function $W_\Lambda[\rho]$
must cancel against the cutoff dependence of the quantum theory
for the soft modes. This constraint can be formulated as a renormalization
group equation for $W_\Lambda[\rho]$ \cite{JKLW2},
to be presented in Sect. 3 below.

\section{Saturation in the classical approximation}

Consider first the simple approximation where the quantum path
integral in eq.~(\ref{2point}) is evaluated in the saddle-point
(or classical)
approximation  $\delta S/\delta A^\mu=0$, and the weight function
is taken simply as a Gaussian (this is the McLerran-Venugopalan
model \cite{MV}, originally formulated for a large nucleus for which
the Gaussian approximation is expected to work better) :
\be\label{FCLAS}
W_\Lambda[\rho]\simeq \exp\left\{-
\frac{1}{2\mu_\Lambda^2}\int d^2x_\perp \,{\rho_a^2(x_\perp)}
\right\}.\ee
Here,  $\mu_\Lambda^2$ is to the total color charge squared (per
unit area) of the partons with $p^+>\Lambda$.

The classical approximation of eq.~(\ref{2point}) reads
(with ${\tilde x}\equiv (x^-,{x}_{\perp})$) :
\be\label{clascorr}
\langle A^\mu(x^+,\tilde x)A^\nu(x^+,\tilde y)\rangle_{cl}\,=\,
\int {\cal D}\rho\,\,W_\Lambda[\rho]\,{\cal A}^\mu({\tilde x})
{\cal A}^\nu({\tilde y})\,,\ee
where ${\cal A}^\mu(\tilde x)$ is the solution to the classical
Yang-Mills equations with source $\rho_a(\tilde x)$,
\be
[D_{\nu}, F^{\nu \mu}]\, =\, g\delta^{\mu +}\delta(x^-) 
\rho_a(x_\perp)\,,
\label{cleq0}
\ee
and is time-independent, like $\rho_a$ itself.
The LC gauge solution can be written as \cite{CGC}:
\beq\label{v}
{\cal A}^+={\cal A}^-=0,\qquad 
{\cal A}^i({\tilde x})=\theta(x^-){\cal A}^i_+(x_\perp)
+\theta(-x^-){\cal A}^i_-(x_\perp),\eeq
with ${\cal A}^i_+(x_\perp)$ and ${\cal A}^i_-(x_\perp)$
related to $\rho(x_\perp)$ via a
non-linear equation. Note that the vector potential
${\cal A}^i(x^-,{x}_{\perp})$ is discontinuous at $x^-=0$,
so the associated electric field ${\cal F}^{+i}\equiv \del^+ {\cal A}^i$
is localized at the light-cone (i.e., within the support of the source):
\be\label{FDELTA}
{\cal F}^{+i}(\tilde x)= \delta(x^-)\Bigl({\cal A}^i_+(x_\perp)
- {\cal A}^i_-(x_\perp)\Bigr).\ee

By using this approximation, let us compute the
gluon distribution function, that is,
the number of gluons per unit 
of $x$ in the hadron wavefunction having transverse momentum less than $Q$.
This is defined as (with $\tilde k^\mu\equiv (k^+,{\bf k}_\perp)$ and 
$k^+=xP^+$) :
\be\label{GDF}
xG(x,Q^2)\,=\,\frac{1}{\pi}\int {d^2k_\perp \over (2 \pi)^2}\,\Theta(Q^2-
k_\perp^2)\,\Bigl\langle F^{+i}_a(x^+,\tilde k)
F^{+i}_a(x^+,-\tilde k)\Bigr\rangle.\ee
In the classical approximation (\ref{FDELTA}),
$F^{i+}(x^+,\tilde k)\approx {\cal F}^{i+}(k_\perp)$ 
is independent of both $x^+$ and $k^+$, and 
(with $\Delta {\cal A}^i\equiv{\cal A}^i_+ - {\cal A}^i_-$, and
the classical average defined as in eq.~(\ref{clascorr})) :
\beq\label{GCL}
x G_{cl}(x,Q^2)&=&\frac{A}{\pi}
\int^{Q^2} {d^2k_\perp \over (2 \pi)^2}
\int d^2x_\perp\,{\rm e}^{-ik_\perp\cdot x_\perp}
\Bigl\langle \Delta {\cal A}^i_a(0_\perp)\,
\Delta {\cal A}^i_a(x_\perp)\Bigr\rangle_{cl}.\eeq
Here, $A$ is the hadron transverse aria, and we have assumed
transverse homogeneity.
Thus, $x G_{cl}(x,Q^2)$ is independent of $x$, which reflects the
absence of quantum evolution in the present, classical approximation.
With the Gaussian weight function (\ref{FCLAS}), and the non-linear
classical solution (\ref{v}), the classical gluon distribution
(\ref{GCL}) can be computed exactly \cite{JKMW97,KM98}. One obtains:
\be\label{SATN}
\Bigl\langle \Delta {\cal A}^i_a(0_\perp)\,
\Delta {\cal A}^i_a(x_\perp)\Bigr\rangle_{cl}
\,=\,\frac{N_c^2-1}{\pi\alpha_s N_c}\,\frac{1-{\rm e}^{-x_\perp^2
\ln(x_\perp^2 \Lambda_{QCD}^2) Q_s^2/4}}
{x_\perp^2}\,,\ee
where $Q_s\propto \alpha_s\mu_\Lambda$ is the {\it saturation momentum},
and is a priori a function of $\Lambda$.
Remarkably, this equation displays
saturation via non-linear effects in the classical solution.
This interpretation can be made sharper by going to momentum space.
If $N(k_\perp)$ is the Fourier transform of (\ref{SATN}),
\be N(k_\perp) \,\propto\, \alpha_s (Q_s^2/k_\perp^2) \quad{\rm for}
\quad k_\perp^2\gg Q_s^2,\ee
which is the normal perturbative behaviour, but
\be N(k_\perp) \,\propto\, {1\over \alpha_s}\,\ln\,
\frac{k_\perp^2}{Q_s^2}\quad{\rm for}\quad k_\perp^2\ll Q_s^2,\ee
which shows a much slower increase, i.e., saturation, at
low momenta (with $k_\perp\gg \Lambda_{QCD}$ though).
According to eq.~(\ref{SATN}), saturation is also a statement
about the maximum field intensity that can be reached in the system:
the classical field never becomes larger than $A^i\sim 1/g$.
This is the maximal occupation number for a classical field, since larger
occupation numbers are blocked by repulsive interactions of the gluon field.  

\section{The non-linear evolution equation}

In writing down the effective theory for soft gluons
in eq.~(\ref{2point}),
we have assumed that the influence of the fast gluons can be
reproduced by a classical color source $\rho$
with a peculiar structure (time-independent, and localized
at $x^-=0$), and some (still unspecified) weight function 
$W_\Lambda[\rho]$.
In this section, we show how to construct this effective 
theory, step by step, by integrating quantum fluctuations 
in successive layers of $p^+$.

To this aim, it is convenient to consider
a sequence of two effective theories (``Theory I''
and ``Theory II'') defined as in eq.~(\ref{2point}),
but with different separation scales: $\Lambda$
in the case of Theory I, and $b\Lambda$ for
Theory II, with $b\ll 1$. That is, Theory II
differs from Theory I in that the
``semi-fast'' fields
with longitudinal momenta between $b\Lambda$ and $\Lambda$ 
have been integrated out, and the associated correlations
have been incorporated at tree-level,
within the new weight function $W_{b\Lambda}$.
The difference $\Delta W
\equiv W_{b\Lambda} - W_\Lambda$ can be obtained \cite{CGC} by
matching calculations of soft ($k^+\ll b\Lambda$)
gluon correlations in both theories. The result can be
expressed as an evolution equation for $W_\tau[\rho]$
(with $\tau\equiv \ln(P^+/\Lambda)$) with respect to variations
in $\tau$.

The quantum corrections due to the semi-fast fields
have to be computed to leading logarithmic
accuracy (LLA), that is, to leading order in 
$\alpha_s\ln(1/b)$---indeed, it is only to this accuracy that the
hierarchy of scales in eq.~(\ref{HL}) is satisfied, and
the matching is possible---,
but to all orders in the classical fields and sources (since,
in the saturation regime of interest, the non-linear effects are
so strong that cannot be expanded in perturbation theory).
This specifies the accuracy of the evolution equations
to be obtained.

To this accuracy, there are only two contributions to $\Delta W$ :
the one- and two-point correlators of the fluctuating colour charge
$\delta\rho_a(x)$ of the semi-fast gluons. Specifically, one obtains
\cite{JKLW2,CGC}
\beq
\langle\delta \rho_a(x)\rangle_\rho&\approx&\alpha_s\log(1/b)\,
\delta(x^-)\,\sigma_a ({x}_\perp),\nn
\langle
\delta \rho_a(x)\delta \rho_b(y)
\rangle_\rho&\approx&\alpha_s\log(1/b)\,
\delta(x^-)\,\chi_{ab}(x_\perp, y_\perp)\,\delta(y^-),\eeq
while all the $n$-point correlators with $n\ge 3$ are
of higher order in $\alpha_s$. In these equations,
$\langle\cdots \rangle_\rho$ denotes a quantum expectation value
over the semi-fast fields at fixed $\rho\,$. Furthermore,
$\sigma_a ({x}_\perp)$ and $\chi_{ab}(x_\perp, y_\perp)$
are generally non-linear functionals of $\rho(x_\perp)$ given by one-loop 
diagrams within Theory I
(with loop momenta restricted to the strip: $b\Lambda < |p^+| < \Lambda$).

In terms of these functions, the evolution equation for $W_\tau[\rho]$
reads \footnote{In condensed notations where, e.g., $\rho_{x}$ stands 
for $\rho_a(x_\perp)$, and repeated indices are understood to
be summed (integrated) over.} \cite{JKLW2}
\be\label{RGE}
{\del W_\tau[\rho] \over {\del \tau}}\,=\,
 \alpha_s \left\{ {1 \over 2} {\delta^2 \over {\delta
\rho_x \delta \rho_y}} [W_\tau\chi_{xy}] - {\delta \over {\delta \rho_x}}
[W_\tau\sigma_x] \right\}\,.
\ee
This functional equation is equivalent to an infinite hierarchy of 
ordinary equations for the correlators of the charge density.
For instance, by multiplying eq.~(\ref{RGE}) with $\rho_x\rho_y$
and functionally integrating over $\rho$, one obtains an evolution
equation for the two-point function:
\be\label{RGE2p}
{d\over {d\tau}}\,
\langle\rho_x\rho_y\rangle_\tau\,=\, \alpha_s\,\langle\chi_{xy}
+\rho_x\sigma_y+\sigma_x\rho_y\rangle_\tau\,,\ee
which in general, however, involves also the higher $n$-point 
functions, via $\sigma$ and $\chi$.
But in the weak source approximation, i.e., with $\sigma$ and $\chi$
computed to lowest order in $\rho$, this becomes a closed equation
for $\langle\rho\rho\rangle$ which has been shown \cite{JKLW1}
to be equivalent to the BFKL equation, as necessary on physical grounds.
This is a highly non-trivial check of the effective theory
in eq.~(\ref{2point}).

In order to study saturation, however, one needs eq.~(\ref{RGE}) 
in the regime of strong background fields and sources (${\cal A}^i\sim 1/g$
and $\rho\sim 1/g^2$; cf. eqs.~(\ref{SATN}) and (\ref{cleq0})), which
requires for an {\it exact} calculation of the coefficients $\sigma$ 
and $\chi$. This has been done recently \cite{CGC}, via a lenthy
calculation which had to cope with difficulties related to gauge-fixing,
the axial poles in the gluon propagator, and the proper definition
of the singular limit $\rho(x) \to \delta(x^-)\rho({x}_\perp)$.
This makes possible to look for solutions to eq.~(\ref{RGE}).
Note that, formally, this is like a functional Schr\"odinger
equation in imaginary ``time'' $\tau$.
An interesting possibility is that the ``Hamiltonian'' in its
r.h.s. has an eigenfunction ${\cal W}[\rho]$ of maximum eigenvalue
$\lambda$. This would lead to an universal behavior of
$W_\tau[\rho]$ in the small-$x$, or high-energy, limit:
$W_\tau[\rho] \longrightarrow_{\tau\to \infty}\, 
{\rm e}^{\tau\lambda}\,{\cal W}[\rho]$, with $\tau$-independent
${\cal W}[\rho]$.

\end{document}